\documentstyle[epsfig,twocolumn,aps]{revtex}
\begin{document}

\title{Fronts dynamics in the presence of spatio-temporal structured
noises}
\author{Miguel A. Santos$^{1}$\cite{mail} and J.M. Sancho$^{2}$}

\address{
$^{1}$Departament d'Enginyeria Mec\`anica, Universitat Rovira i
Virgili,
Ctra. Salou s/n, 43006 Tarragona. Spain}

\address{$^{2}$
Departament d'Estructura i Constituents de la Mat\`eria,
Universitat de Barcelona,
Av. Diagonal 647, 08028 Barcelona. Spain.
 }

\date{\today }

\maketitle
\begin{abstract}
Front dynamics modeled by a reaction-diffusion equation are studied under
the influence of spatio-temporal structured noises.
An effective deterministic model is analytical derived where the noise
parameters, intensity, correlation time and correlation length appear
explicitely. 
The different effects of these parameters are discussed for 
the  Ginzburg-Landau  and  Schl\"oegl models.
We obtain an analytical expression for the
front velocity as a function of the noise parameters. Numerical
simulations results are in  a good agreement with the theoretical
predictions.
\vspace{8pt}

\vspace{8pt}

\end{abstract}


\section{Introduction}
The role of external fluctuations in extended systems is a subject of very
active research because of its relevance in pattern formation in
non-equilibrium systems \cite{mikhailov,JgoJmbook}.
A simple example of a dynamical
pattern is a front moving at constant velocity. Fronts can be easily
modelized by a reaction-diffusion equation with two steady states of
different stability \cite{cross}. The study of the front dynamics under
the influence of noises is relevant not only from theoretical point of
view \cite{mikhailov2,engel,schymanski,pasquale,J96,JLJm98} but also from
practical point as recent
works on chemical kinetics have shown
\cite{irene1,showalter1,showalter2,irene2}. In these experiments a
chemical wave moves under the influence of and external fluctuating
illumination 
which is projected in the reactive medium. This external source of noise
has finite intensity, correlation time and correlation length. Thus, in
this system we face with spatio-temporal structured noise and not with
a white noise. The present work could be useful to clarify the role
of these parameters on propagating structures as studied experimentally
in \cite{irene1}.

Previous studies dealt with this problem under the simplified assumption
of
white ($\delta$-correlated) external fluctuations
\cite{mikhailov2,engel,schymanski,pasquale,J96,JLJm98}. Nevertheless one
can ask about the
correctness of this assumption to modelize real noises. To answer this
question, at least it would be necessary to calculate the first
contributions of
the finite value  external noise parameters. If this corrections are
controlled then one can get confidence on the simplified assumption of
white noise.  

Since the early work of {\sl Schl\"ogl} \cite{schloegl} on the effects
of fluctuations on a chemical interface, an intensive work has been
devoted to describe the related problem of front propagation
in the presence of an external noise source
\cite{mikhailov2,engel,schymanski,pasquale,J96,JLJm98}. 
A complete study for the case of a white noise was presented in
\cite{J96,JLJm98} in which the front velocity and its diffusive dispersion
behavior was computed in terms of the effective white noise intensity.
It was found that the
velocity of the front increases with the noise intensity due to a
systematic contribution to the kinetic terms. 
Actually the dispersion is subdiffusive for the so called
{\sl pulled} fronts, as recently has been shown \cite{Andrea}.
A variety of approaches has been adopted, from projection techniques
\cite{schymanski,pasquale} to scaling arguments \cite{Andrea}, including a 
stochastic
version of the multiple scale analysis as well as a
non-systematic noise expansion \cite{JLJm98}.
The last one may be the most simplified one which grasps the systematic
contribution of the noise to the dynamics of the system. This contribution
usually appears as a renormalization of the reaction parameters and it
is the origin of the well known shift of the front velocity.

The general aim of this paper is to find what are the most relevant
effects of a real noise on two different models which exhibit front
propagation. We will see that for a fixed noise intensity, the
noise correlation time is a relevant parameter which interpolates the
results of the white noise limit with those of the deterministic case, but
the role of the correlation length is different.
 
Here, we will derive an analytical expression of the effect of a
spatio-temporal structured noise on an extended system governed by a 
Langevin reaction-diffusion equation with multiplicative noise. 
We will closely follow the guidelines settled in \cite{SSKG} for a one 
variable system, and also those in \cite{LJm88} for multivariable system
to deal with non white noises in extended systems.
Our main difference is that our analysis is done in the continuum space, 
and also that we present a more simplified way to get the first order 
contribution of the noise in the correlation time $\tau$ which avoid 
the integration of a response function. Here we note that the 
continuum Langevin description may perfectly be adequate for 
describing reactive
fronts even though chemical systems are discrete in nature, as was
shown in \cite{Lema}. 

Our theoretical predictions have been applied to two
systems: 
the Ginzburg-Landau and  the Schl\"ogl  models. 
In the first case the noise induces the front by controlling the stability
of the new state versus the other unstable 
steady state. In the second case, the noise does not change the steady
states but controls its dynamics.

The outline of the paper is as follows.
Section II contains  the main theoretical results and a discussion of some
limiting cases. There we present the derivation of an  effective dynamical
equation
which grasps the systematic contribution of the different  noise
parameters.
In section III, we apply these results to obtain explicit predictions for 
the two  models already mentioned and we discuss the numerical 
results obtained
for these them and their comparison with the analytical predictions.
In section IV we summarize our conclusions.
Several Appendices are devoted to 
technical aspects of our analytical methodology and the implementation of
a particular algorithm to generate a patio-temporal structured noise.


\section{Effective dynamical model}

We consider the following stochastic partial differential equation as a
representative description of reaction-diffusion systems under
multiplicative fluctuations: 
\begin{eqnarray}
\frac{\partial \psi(x,t)}{\partial t} &=& 
{\cal L} \lgroup \psi(x,t),\partial_x,a\rgroup 
+ \epsilon^{1/2} g(\psi) \eta(x,t) 
\label{SPDE} .
\end{eqnarray}
where ${\cal L}$ is a reaction-diffusion operator which explicitely reads,
\begin{equation}
{\cal L} \lgroup \psi(x,t),\partial_x,a\rgroup\,=\, 
D\frac{\partial^2 \psi(x,t)}{\partial x^2} + f(\psi,a). 
\label{nldo}
\end{equation}
$f(\psi,a)$ and $g(\psi)$ are the reaction term and 
the coupling term with external fluctuations 
respectively, and
$\eta(x,t)$ is a Gaussian spatio-temporal structured noise with the
following statistical properties,
\begin{eqnarray}
\langle \eta(x,t) \rangle &=& 0 \nonumber\\
\langle \eta(x',t') \eta(x,t) \rangle &=& 
G(|x-x'|,|t-t'|) = \nonumber\\
&=&C(|x-x'|) \gamma(|t-t'|) .
\label{etastat}
\end{eqnarray}
Also, for simplicity but not strictly necessary, we have made the
assumption that this correlation function factorizes in a spatial
and temporal part. To fix the notation, and following the commonly 
accepted generic prescription \cite{JgoJmbook},
we define the three parameters of the noise, intensity, correlation time
and correlation length,  as follows:
\begin{eqnarray}
\sigma^2&\equiv&\int_0^{\infty}\,ds\,\int_{{\cal R}}\,dr\,G(r,s) \nonumber
\\
\tau &\equiv& \frac{1}{\sigma^2}\,
\int_0^{\infty}\,ds \,\int_{{\cal R}}\,dr\,G(r,s)\,s \label{etaparam}\\
\lambda^2  &\equiv& \frac{1}{\sigma^2}\,
\int_0^{\infty}\,ds \,\int_{{\cal R}}\,dr\,G(r,s)\,r^2  \,.\nonumber
\end{eqnarray}

We pursue here to find the systematic and most relevant effects of
this type of noise. In general, the noise has two important effects, 
systematic  and fluctuating ones, which cannot be exclusively associated
with
the deterministic and stochastic terms of Eq. (\ref{SPDE})
respectively.
In fact noise acts in
two different scales \cite{JLJm98}. 
Fast fluctuations in a short time scale modify the
front shape and thus producing an effective front with different deterministic
properties. On the other hand, the slow fluctuations are responsible of
the diffusive dispersion of the front position .   

A naive way to get  these systematic effects of the
fluctuations is by analyzing the noise term in (\ref{SPDE}).
Due to de multiplicative character of the noise,
although $\eta$ has zero mean, this is not the
case for this stochastic term,
\begin{equation}
\epsilon^{1/2}\langle g\lgroup\psi(x,t)\rgroup \eta(x,t)\rangle \equiv
\langle\Phi\lgroup\{\psi\}\rgroup\rangle \neq 0  .
\label{sigmadef}
\end{equation}
which, as a
consequence, will give a
net contribution to the dynamics. 
This can be explicitly shown in the following way.
By adding and substructing 
$\Phi\lgroup\{\psi\}\rgroup$  to our original dynamical 
equation (\ref{SPDE}), 
we can write this equation as
\begin{eqnarray}
\frac{\partial \psi(x,t)}{\partial t} =
{\cal L} \lgroup \psi(x,t),\partial_x,a\rgroup +
\Phi\lgroup\{\psi\}\rgroup 
+\,\epsilon^{1/2}\,{\cal R}\lgroup\psi,x,t\rgroup,
\nonumber\\
\epsilon^{1/2}\,{\cal R}\lgroup\psi,x,t\rgroup \equiv \epsilon^{1/2} g(\psi)
\eta(x,t) - \Phi\lgroup\{\psi\}\rgroup.
\label{SPDE2}
\end{eqnarray}
This dynamics is statistically equivalent to the original one.
Note that for the new noise term it is 
$\langle{\cal R}\lgroup\psi,x,t\rgroup\rangle\equiv 0$
and it has a correlation which can be developed in powers
of $\epsilon^{1/2}$.
We make now the Ansatz that if the noise allows for a {\sl well definite
front structure}, its systematic behavior will be described
by the {\sl deterministic} equation 
\begin{equation}
\frac{\partial \psi(x,t)}{\partial t} =
{\cal L} \lgroup \psi(x,t),\partial_x,a\rgroup +
\Phi\lgroup\{\psi\}\rgroup
\label{sysdyn}
\end{equation}
called the {\sl effective dynamics}. 
In appendix \ref{rspfA} and \ref{rspfB} we present a detailed
calculation of $\Phi$ for {\sl small} $\tau$, which is given by
\begin{equation}
\Phi\lgroup\{\psi\}\rgroup = \epsilon\,D\,C(0)\,\tau\,
g'\,g''\,\Big(\frac{\partial\psi (x,t)}{\partial x}\Big)^2\,
+\,h\lgroup\psi(x,t)\rgroup\,
\end{equation}
where
\begin{eqnarray}
h\lgroup\psi (x,t)\rgroup\,\equiv\,
f\lgroup\psi (x,t),a\rgroup\,
+\,\nonumber \\
\epsilon\,\{C(0)\,+\,D\,C''(0)\,\tau\}\,
g\lgroup\psi(x,t)\rgroup g'\lgroup\psi (x,t)\rgroup\,\nonumber  \\
-\,\epsilon\,C(0)\,\tau\,
g'\lgroup\psi (x,t)\rgroup\,\{ g\lgroup\psi (x,t)\rgroup,f\lgroup\psi (x,t)\rgroup \}
\label{effectivepot} 
\end{eqnarray} 
is  the new effective reaction term. The brackets are defined as
\begin{equation}
\{\,g\,,\,f\,\}\,\equiv\,g'\,f\,\,-\,\,g\,f' \,,
\end{equation}
and the primes on $f(\psi)$ and $g(\psi)$ indicate the derivative with 
respect to $\psi$.

Thus, we have ended up with one of the most important 
results of this paper, Eq.(\ref{sysdyn}), which contains 
the systematic contribution of the noise to
our original dynamics (\ref{SPDE}) up to first order in $\tau$. 
In this paper will not study 
the effect of $\cal R$.  
This term is only relevant for those non
systematic effects of the noise such as, for example, the dispersion of
the front. 
The dependence on the parameter $\lambda$ is included in $C(0)
\sim \sigma^2 \lambda^{-1}$
and $C''(0) \sim \sigma^2 \lambda^{-3}$. 
As will be seen below, $C(0)$ is the most relevant quantity.
Thus the
main effect of the correlation length through $C(0)$ is trivial. 
For this reason we will pay
more attention to the non-trivial influence of the term $C''(0)$ fixing
$C(0)$ independent of $\lambda$.
   
A first check of the previous results  will be provided by 
considering the better known case of temporal white noise in a lattice.
Here we will first 
define the proper limit by which (\ref{etastat}) becomes a temporal white
noise, 
and then see if (\ref{sysdyn}) correctly reproduces the results found 
in Ref. \cite{J96}.

The temporal white noise in time has a correlation, 
\begin{equation}
\langle \eta(x',t') \eta(x,t) \rangle = 
2\,C(x-x')\,\delta(t-t')\, \label{whitestat},
\end{equation}
where the spatial white noise limit is given by
\begin{equation}
\lim_{\lambda\to 0}\,C(x-x')\,=\,
\delta (x-x'). 
\end{equation}
In a one dimensional lattice this takes the form of 
\begin{equation}
\lim_{\lambda \to 0}\,C_{ij}
\,\equiv\,\frac{\delta_{ij}}{\Delta x}\,.
\end{equation}
In this limit  one can see that all the integrals in
(\ref{gpartial}) vanish
except the first one.
The systematic
dynamics of (\ref{sysdyn}) is then that of 
the effective reaction term given now by ($\sigma^2=1$)
\begin{equation}
h_{\eta}\lgroup\psi (x,t)\rgroup\,\equiv\,
f\lgroup\psi (x,t),a\rgroup\,
+\,\epsilon(0)\,
g\lgroup\psi(x,t)\rgroup g'\lgroup\psi (x,t)\rgroup, \label{wneffectivepot}
\end{equation}
with $\epsilon(0)\,\equiv\,\epsilon\,C(0)=\epsilon/{\Delta x}$. 

In this way we have recovered the results found in Ref. \cite{J96}. 
Note that one cannot consider right from the beginning a white noise in 
space because the ill defined  $\delta(0)$.

The value of $C''(0)$ for a spatial white noise in the lattice is
evaluated as,
\begin{equation}
C''(0)\,=\,\frac{C(1)-2C(0)+C(-1)}{(\Delta x)^2}\,=\,
-\,\frac{2}{(\Delta x)^3}, 
\label{wnprescription}
\end{equation}
where $C(\pm 1)=0$ has been used.
For the case of a  spatial structured noise with $\lambda$ finite,
all the integrals in (\ref{gpartial}) can be evaluated.

\section{Applications and Numerical results}

We will now study the effects of a colored noise 
for two particular types of
couplings $g(\phi)$: a linear and a nonlinear one, which correspond to
the Ginzburg-Landau  and Schl\"{o}gl models respectively. 
The noise will enter in the standard way \cite{HL85}
as small fluctuations of the control parameter $a$,
\begin{equation}
a\,\to\,a\,+\,\epsilon^{1/2}\,\eta(x,t),\label{stndextn}
\end{equation}
and thus the Langevin-type coupling function is given by,
\begin{equation}
g\lgroup\psi (x,t)\rgroup\,=\,
\frac{\partial f\lgroup\psi ,a\rgroup}{\partial a}\, \label{extnclp}.
\end{equation}

Numerical simulations of the Eq.(\ref{SPDE}) for the different models have
been performed in a one-dimensional lattice of mesh-size $\Delta x=.5$.
The length of the system is $L=600$.
We have used a Heun algorithm \cite{JgoJmbook} with a time step $\Delta
t=.01$.
In all cases, $D=1$ and $a=-.1$, except for the Schl\"{o}gl
model, where different values of $a$ have been used.

The noise is generated with a spatial and temporal structure
as a Gaussian random numbers at each lattice point. The correlation
function factorizes as in Eq.(\ref{etastat}).
The temporal part has a exponential decay 
(Ornstein-Uhlenbeck process) with a correlation time $\tau$, while the
spatial correlations have
a triangular shape with a  correlation
length $\lambda$. The numerical implementation 
of such a
noise is described in appendix \ref{stcn}.
 
The initial condition  for 
the Ginzburg-Landau model is a small pulse with a height of $0.01$ and
located
at the middle of the spatial domain. 
In this way the initial perturbation
will spread off as two fronts propagating in opposite directions.
For the Schl\"{o}gl model the initial field is a step-like
function of value $\psi(0,0)=1$  and $\psi(x,0)=0$ in the rest of the
spatial domain.
The numerical calculation  of the mean front velocity and
the steady state behind the front have been done as in Ref. \cite{MaJm99}.

\subsection{Linear coupling: The  Ginzburg-Landau model}

This model have already been considered
in the context of noise-induced fronts \cite{MaJm99}. 
We will now study how that 
picture is modified by a  spatio-temporal structured  noise. 
For this model the kinetic term is, 
\begin{equation}
f\lgroup\psi ,a\rgroup\,=\,
-\,\psi\,\lgroup a\,+\,\psi^2\rgroup\,, \label{RGL}
\end{equation}
and as a consequence, the noise coupling term  is linear,
\begin{equation}
g(\phi) = \phi.
\label{lcp}
\end{equation}.

We will have then that the
effective dynamics  given
by Eq. (\ref{sysdyn}) is
\begin{equation}
\frac{\partial \psi(x,t)}{\partial t} = D\frac{\partial^2 \psi(x,t)}{\partial x^2}\,
+\,h\lgroup\psi(x,t)\rgroup\,\label{RGLsysdyn}
\end{equation}
with a new kinetic term (\ref{effectivepot}) given now as,
\begin{equation}
h\lgroup\psi (x,t)\rgroup\,=\,
-\,\psi\,\lgroup\,a'\,
+\,b'\,\psi^2\rgroup\,, \label{RGLeffectivepot}
\end{equation} 
where the new kinetic parameters are,
\begin{eqnarray}
a'&=&a\,-\,\epsilon\,
\{C(0)\,+\,D\,C''(0)\,\tau\} \nonumber \\
b'&=&1\,+\,2\,\epsilon\,C(0)\,\tau. \label{RGLrnpar}
\end{eqnarray}

Following the linear marginal stability criteria \cite{WvS89}, the velocity 
of a {\sl pulled} front is controlled by the linear term as,
\begin{eqnarray}
v^*_l\,&=&\,2\,\sqrt{D\,(-a')}  \nonumber \\
&=&\,2\,\sqrt{D\,\lgroup-a\,+\,\epsilon (0)\,(1\,+\,
\frac{D\,C''(0)}{C(0)}\,\tau)\rgroup} \label{GLv}.
\end{eqnarray}
Note that $a'< 0$ in order to have a front. 
This result, however, has been deduced  for a small enough
$\tau$. Nevertheless, we can conjecture 
a generalization of  Eq.(\ref{GLv}) for any value of $\tau$
considering that the values of the velocity for $\tau=0$ (temporal white
noise limit) and for $\tau=\infty$ (deterministic case) are known.
In this way, the simplest {\em regularization} of Eq.(\ref{GLv}), which 
is  a monotonous function on $\tau$, is
\begin{equation}
v^*_{\tau}\,=\,2\,\sqrt{D\,\lgroup-a\,+\,\frac{\epsilon (0)}{1\,-\,
\frac{D\,C''(0)}{C(0)}\,\tau}\rgroup}. \label{GLvUCNA} 
\end{equation}
Moreover,  Eq.(\ref{GLvUCNA}) generalizes the temporal white noise result
in terms on a renormalized noise intensity defined as, 
\begin{equation}
\epsilon_{\cal R}\equiv \frac{\epsilon (0)}
{1\,-\,\frac{D\,C''(0)}{C(0)}\,\tau},
\label{e_R}
\end{equation}
which does not present any singularity because
always $C''(0) < 0 $. 

Taking $\epsilon(0)=\epsilon C(0)$ as a constant we have the following 
behavior. For increasing $\tau$ ($\lambda$ fixed), $\epsilon_{\cal
R}$ decreases, and we arrive up to the deterministic value of $v$.
Nevertheless for fixed
$\tau$ and increasing $\lambda$, $C''(0)\tau / C(0) \sim \tau/\lambda^2
$, and then $\epsilon_{\cal R}$ increases, and as a consequence, $v$
increases up to the temporal white noise limit. This is a non trivial
effect of $\lambda$ which needs a finite value of $\tau$ to appear. 

As already discussed in \cite{MaJm99}, another important quantity in this
model is the field behind the
front which is induced by the noise, and thus it is highly
fluctuating. From Eq.(\ref{RGLeffectivepot}), the homogeneous
deterministic 
stationary value behind the front can be calculated as,
\begin{equation}
 \psi_{st} =  \sqrt{\frac{-a'}{b'}}. \label{GLstpsi}
\end{equation}

As we don't know how the higher order corrections on
$b'$ are, we 
expect a poorer agreement for $\psi_{st}$ 
than for the velocity.
However, we can get an idea of the relevance of  $b'$
by numerically inspecting the quotient of $v^*_{\tau}$ over $\psi_{st}$
which depends on $b'$. Indeed, 
from Eqs. (\ref{GLv}) and (\ref{GLstpsi}), this dependence is
\begin{equation}
\left(b'\right)^{1/2}\,=\,
\frac{1}{2}\,\frac{v^*_{\tau}}{\psi_{st}\,\sqrt{D}}.
\label{GLb'}
\end{equation}

The analytical predictions (\ref{GLvUCNA}), (\ref{GLstpsi}) and
(\ref{GLb'}) are important results of this paper that 
will be checked numerically.
Due to the different role of the noise parameters $\tau$ and $\lambda$ we
will discuss two cases separately.
 
\subsubsection{Spatial white noise in the lattice}

\begin{figure}
\centerline{\psfig{figure=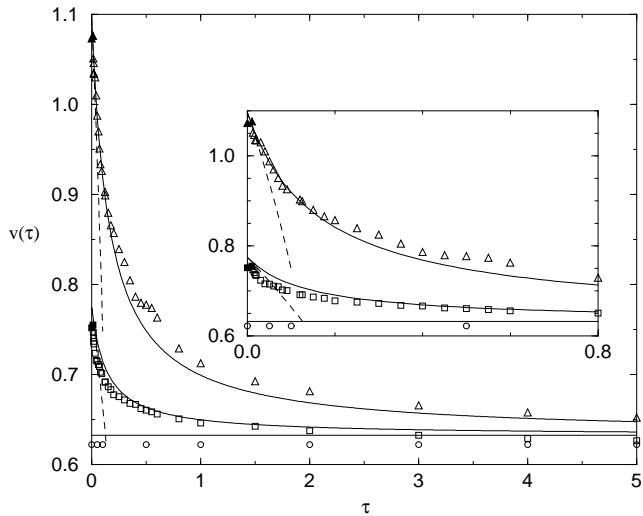,width=8.5cm}}
\caption{\em
Front mean velocity versus noise corelation time $\tau$ for the
Ginzburg-Landau
model in the presence of a Ornstein-Uhlenbeck noise in time,
white in space.
Values of the parameters:
Circles correspond to the deterministic
case, $v^*_{sim}=.62$; Triangles, to $\epsilon(0)=.2$,
and squares, to $\epsilon(0)=.05$; dashed lines are the analytical
prediction up
to $O(\tau)$ (\protect\ref{GLv}), whereas continuous lines represent the
corrected  prediction  (\protect\ref{GLvUCNA}).
See text for the values of the other parameters. \label{vtau}
}
\end{figure}

For this case  all the simulations
perfectly agree with the theoretical results of Eq. (\ref{GLvUCNA}). In
fact, fixing
the
noise intensity $\epsilon(0)$ and increasing $\tau$ the
mean velocity of the front drops monotonously to the deterministic
value (see Fig. \ref{vtau}. All figures are in dimensionless units).

Our analytical calculation (dashed lines) only describe the
corrections to the white noise case at order $O(\tau)$. However,
this can be considered quite relevant as the dependence of
$v$ and $\psi_{st}$ on
$\tau$ drops down very rapidly near $\tau=0$, and our first
order approximation succeeds to grasps this pronounced slope
(see inset of fig.\ref{vtau}).
Moreover, the extended analytical prediction (\ref{GLvUCNA}) shows a very
good agreement with numerical data for all values of $\tau$.

\begin{figure}
\centerline{\psfig{figure=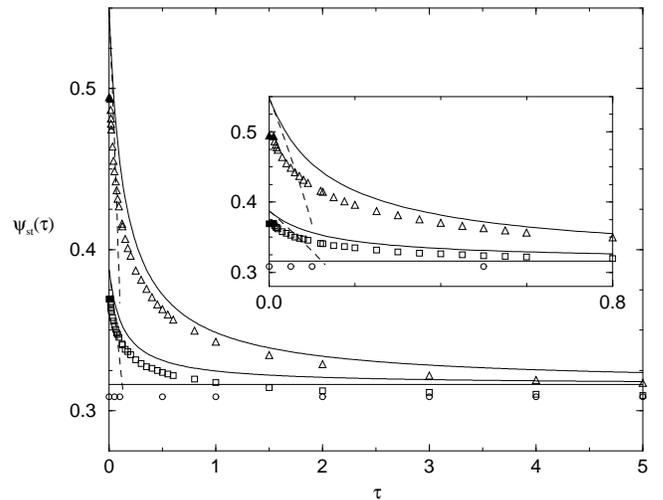,width=8.5cm}}
\caption{\em
Mean stationary value of the field versus $\tau$.
The inset is a amplification of the small $\tau$ domain.
See previous figure for the symbol notation.}
 \label{phi-1}
\end{figure}

\begin{figure}
\centerline{\psfig{figure=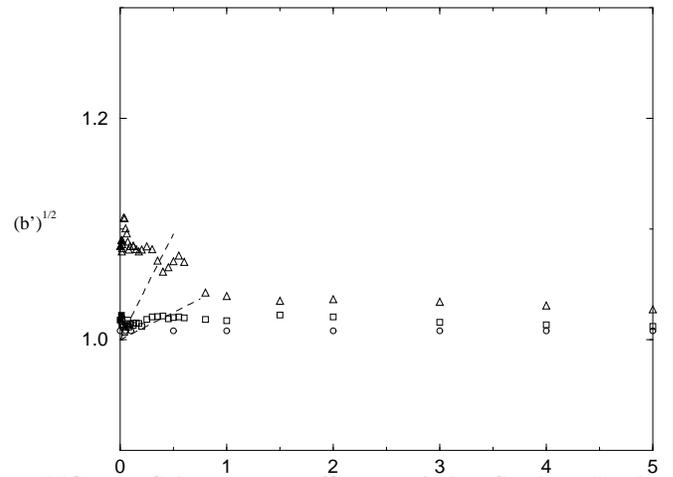,width=8.5cm}}
\caption{\em
Cubic term coefficient of the Ginzburg-Landau model as
the quotient of $v$ to $\phi$ versus $\tau$.  Notation
is the same as in previous figures.\label{q-1}
}
\end{figure}

With respect to the average mean field, the
agreement is more qualitative (see Fig.\ref{phi-1}).
The numerical results for the effective parameter $b'$, evaluated from Eq.
(\ref{GLb'}),  are represented in
Fig.\ref{q-1}.
They support the
initial growth of $b'$ predicted by the theory. 
For greater
values of the correlation time  $b'$ stays bounded by its deterministic
value $b'_d=1$.
This fact may explain why numerical values
of $\psi_{st}$ seem to depend only  on  the linear
coefficient $a'$.

From previous work \cite{MaJm99}, we already know that
$\psi_{st}$ has a systematic
error that slightly increases with noise intensity.
Hence, the deviation found here (Fig.\ref{q-1}) is not due to
the presence of temporal correlations, but a problem related to
the fact that we are measuring a highly fluctuating quantity as it is
$\psi_{st}$. In any case,
our theoretical prediction is consistent qualitatively with numerical
simulation results.
  
\subsubsection{Spatio-temporal structured noise}

To study the non trivial effects of a finite correlation length
on the dynamics of the front, we have to pay attention to the effects
coming from the quantity $C''(0) \sim \sigma^2 / \lambda^{-3}$. 

\begin{figure}
\centerline{\psfig{figure=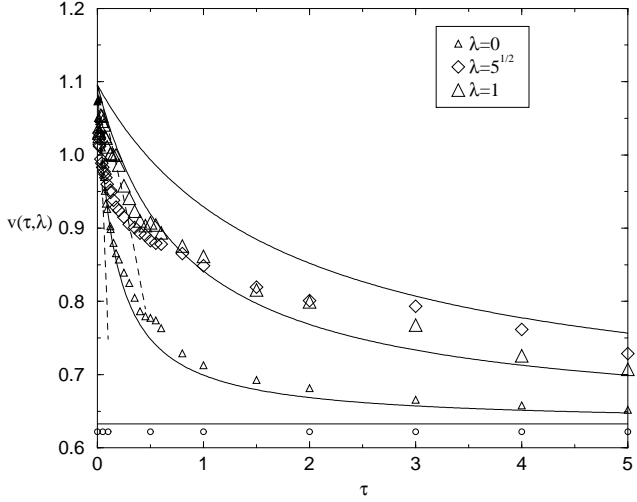,width=8.5cm}}
\caption{\em
Front mean velocity versus $\tau$ for different correlation
lengths. Here it is $\epsilon(0)=.2$.
\label{vs-1}
}
\end{figure}

\begin{figure}
\centerline{\psfig{figure=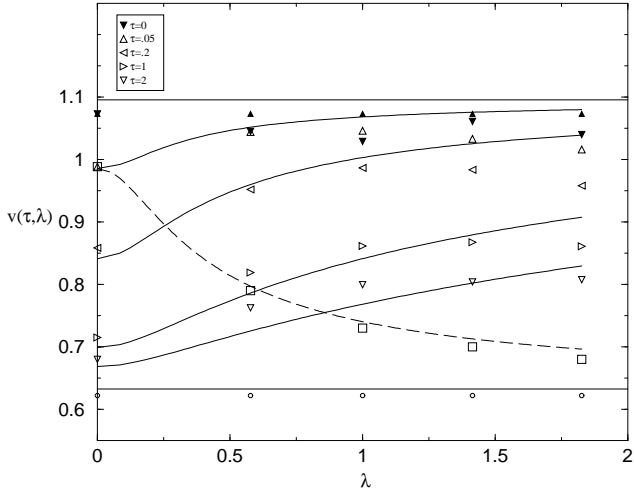,width=8.5cm}}
\caption{\em
Front mean velocity vs. $\lambda$ for different correlation
times. $\epsilon(0)=.2$ for hollow triangles.
Filled up-triangles correspond to the white
noise case, while circles to the deterministic case.
Squares correspond to the trivial noise effects with fixed noise
intensity for $\tau=.05$ and the
long-dashed line is its theoretical prediction (\protect\ref{GLvUCNA})
(see text).
Each set of vertical points along the horizontal axes correspond,
in increasing order of $\lambda$,
to $m\,=\,0,1,2,3,4$, respectively. See appendix \protect\ref{stcn}.\label{v2s}
}
\end{figure}

In Fig.(\ref{vs-1}) we can see the front velocity versus $\tau$ for
different values of the correlation length $\lambda$ of the
noise. Continuous lines correspond to the
analytical prediction (\ref{GLvUCNA}). As can be seen, for
a finite correlation length the agreement is only qualitative, and
improves
for noises not too much away from a spatial white noise (triangles).
This behavior can better be appreciated  in Fig.(\ref{v2s}) where we
have plotted the velocity versus $\lambda$ for different values
of the correlation time.

According to our definition of $\lambda$ (\ref{definitions}), we have that
$\lambda=0$ for the
spatial white noise in the lattice.
Here it can be seen that our analytical scheme may qualitatively
describe the effects of a finite $\lambda$ only for
small values of $\tau$ and  $\lambda$, which is not the case for
$\lambda= 0$
where the accordance is very good.
For a finite $\lambda$, we have also observed a clear
departure from
the analytical results in the case of
the dependence on $\tau$ of the mean stationary value of the field
behind the front, as well as for the cubic coefficient. 
Also, the numerical results
show a systematic decrease of the velocity when increasing the
correlation length for for $\tau=0$, which is not predicted by 
our analytical results (\ref{GLvUCNA}).

On the other hand, for $\tau>0$, the velocity tends to grow with
$\lambda$ for small correlation lengths as the theory predicts.
Indeed, this agrees with what  we have observed in preliminary 
numerical simulations
for a quenched white noise. Hence, for increasing $\lambda$ ($\tau$ fixed)
the numerical results suggests a non monotonous behavior of the velocity,
which may increase at small $\lambda$, but would always decrease at
long correlation lengths.
  
We have not found yet an explanation for this effect.
We believe that 
there is a interplay between the correlation length of the noise
and the typical length of the front, which is given by its width.
Indeed, this effect could be related to the observed distorsion of 
the leading edge of the front and the possible formation of a prefront
in the presence of a large spatial correlation length
of the noise. Then our initial assumption of a well defined 
mean front profile is
not fulfilled and, as a consequence, the theoretical scheme cannot be
applied.
  
For the sake of completness, in this figure
we also show what are the trivial effects of
of a finite correlation length. 
In this case, the noise has been generated such that the 
noise intensity $\sigma^2$, defined in (\ref{etaparam}), remains constant. 
This can be
acomplished by changing the previous wheighting function $g_i$
by a factor $(\sqrt{2\,m\,+\,1})^{-1}$ (See appendix \ref{stcn}).
Squares correspond to the numerical results ($L=2400$). The dashed line
is the theoretical prediction (\ref{GLvUCNA}) for this case. 
As can be seen, there is a monotonous decay of the velocity 
because now the most dominant term $\epsilon(0)\sim\sigma^2/\lambda$
also decays with $\lambda$. Thus, the qualitative behavior 
is completely different.
This confirms that the previous studied dependence on $\lambda$
correspond indeed 
to a {\sl non trivial} effect of the correlation length.
Note that our theoretical scheme succeeds better in describing 
quantitatively the trivial effects
of a finite $\lambda$.
  
\subsection{Nonlinear coupling: The Schl$\ddot{o}$gl model}

The general model was introduced by {\sl Schl\"ogl} in \cite{schloegl} 
when studying
the fluctuations of an interface. Here we will consider a particular
version of it that was  studied in the
presence of an external white noise in Ref.
\cite{J96}.
It corresponds to the reaction term, 
\begin{equation}
f\lgroup\psi ,a\rgroup\,=\,
-\,\psi\,\lgroup\psi\,+\,a\rgroup\,
\lgroup\psi\,-\,1\rgroup \label{schloegl},
\end{equation}
which implies a
 nonlinear coupling with the noise,
\begin{equation}
g\lgroup\psi (x,t)\rgroup\,=\,\psi (\psi\,-\,1)\label{nlcp}.
\end{equation} 

Taking into account these definitions, the effective deterministic part of
Eq.(\ref{sysdyn}) becomes,
\begin{eqnarray}
\frac{\partial \psi(x,t)}{\partial t} = D\frac{\partial^2 \psi(x,t)}
{\partial x^2}\,
+\,h\lgroup\psi(x,t)\rgroup\,+ \nonumber \\ 
+\,D\,\epsilon(0)\,\tau\,
2\,(2\,\psi\,-\,1)\,
\Big(\frac{\partial\psi (x,t)}{\partial x}\Big)^2\,
\label{schlglsysdyn}
\end{eqnarray}
where the reaction term is,
\begin{equation}
h\lgroup\psi (x,t)\rgroup\,=\,
a'\,\psi\,+\,b'\,\psi^2\,+\,c'\,\psi^3\,
+\,5\,d'\,\psi^4\,
-\,2\,d'\,\psi^5, \label{schlgleffectivepot}
\end{equation}  
with the effective kinetic parameters,
\begin{eqnarray}
a'&=&a\,+\,\epsilon_{\cal R}\nonumber \\
b'&=&1\,-\,a\,+\,d'\,
-\,3\,\epsilon_{\cal R}\nonumber \\
c'&=&-1\,-\,4\,d'\,
+\,2\,\epsilon_{\cal R}\nonumber \\
d'&=&\epsilon (0)\,\tau, \label{schlglrnpar}
\end{eqnarray}
where $\epsilon_{\cal R}$ was defined in Eq.(\ref{e_R}).

Our point of interest in this model is the mean front velocity because 
the steady states for the front $\psi=0,1$ are not modified
by the noise.
Due to the prefactor  of the KPZ-like term,
for $\psi> 1/2$, any deviation from the homogeneous state $\psi=1$
tends to grow,
while this is opposite at points where $\psi< 1/2$ for any
deviation from the state $\psi=0$. 
Thus, the effect of this term is to shorten the 
width of the front, i.e., to select a greater decay mode of the
front, thus slowing down its propagation 
\cite{WvS89}. 
Hence, the 
expected slowing down of the front due to temporal correlations 
of the noise arises here in two ways: The first one is 
by means of the usual renormalization of the 
coefficients of the reaction term $f(\psi)$. 
The second type of corrections come 
from the new KPZ-like term. 

We expect that this front will exhibit the general regimes of 
front propagation\cite{WvS89}, i.e., a linear ({\sl pulled} front), 
a non-linear and a metastable (both, {\sl pushed} front) regimes. 
In the linear regime, the velocity depends only on the linear
coefficient of the reaction term $a'$, and it is then given by Eq.
(\ref{GLvUCNA}).

While a crossover from a metastable to a nonlinear regime is trivial
to determine, being nothing more than condition $a'=0$,
the transition between the linear and the nonlinear regimes is far more 
complicated to locate. This calculation of this point  requires an
analysis 
of (all) higher power terms of the reaction,  determining a
complete solution in the comoving system and requiring then 
that the asymptotic behavior is such that the coefficient of the slowest
decay mode vanishes.

For the case of a temporal white noise,
as was found in \cite{J96,JLJm98}, there is only a 
renormalization of the parameters of the kinetic terms, in such a way
that the effective dynamics is equivalent to the deterministic
one up to a rescaling of the coefficients. Hence, the location
of the different regimes can be directly determined from those of the
deterministic case.
Unfortunately, to our knowledge, the first procedure is hopeless for 
Eq.(\ref{schlglsysdyn}). Neither can this dynamic directly be compared
with the deterministic case. 
However, the transition between linear and non linear regime is always 
continuous and, as we can correctly describe
the linear regime, this fact will help us to numerically locate the transition 
for this model.

Nevertheless, there is still some hope for an analytical prediction.
The type of dynamics given 
by Eqs. (\ref{schlglsysdyn})-(\ref{schlglrnpar})
usually are relevant near the transition point
$a'=0$,
where the dynamics given by  Eq.(\ref{nldo}) can be simplified by means
of an amplitude expansion. In this case, and as long as 
the noise intensity is low enough,
our effective equation would also lie near threshold ($a' \sim 0$). 
Assuming this situation, the spatial variations of the field take place
on a typical length scale of order $q_o^{-1}\equiv \sqrt{D/a}$.
A crossover between non linear and linear regime means that the
nonlinear terms start to dominate the growth rate off the initial
steady state. Thus, this transition takes place when both, $a'\,\psi$ and
$b'\,\psi^2$ are of the same order of magnitude.
This will be the case for
\begin{equation}
\psi \sim b' \, ;\qquad
b'^2 \sim a'. \label{psimag}
\end{equation}

Then the KPZ and the $\psi^4$ terms, both will be of order $q_o^4$,
while the
term $\psi\,\Big(\frac{\partial\psi (x,t)}{\partial x}\Big)^2$ will be
of order $q_o^5$. Hence, near threshold, only the first three terms of
the effective reaction (\ref{schlgleffectivepot}) will be relevant.
But this equation is just the standard Schl\"{o}gl model that is
exactly solvable. This will have a sense only if it is $c'<0$, which
we will assume to be so due to its expression in (\ref{schlglrnpar}). As
stated above, we will also require $b'>0$.
The new stationary states of this approximation are given by 
\begin{equation}
\psi_{\pm}\,\equiv\,\Big(\frac{-b'}{2\,c'}\Big)\,
\{
1\,\pm\,\sqrt{1\,-\,\frac{4\,a'\,c'}{{b'}^2}}
\}.
\end{equation}
For $a'\,c'<0$ and $b'>0$, it is $\psi_+>\psi_-$.
If we write then the reaction term as
\begin{equation}
h_{cr}\lgroup\psi (x,t)\rgroup\,=\,
c'\,\psi\,(\psi\,-\,\psi_+)\,(\psi\,-\,\psi_-), \label{schlgleffectivepotcr}
\end{equation}
the velocity of a front connecting $\psi=\psi_+$
and $\psi=0$ is given by \cite{VRCY87}
\begin{equation}
v_{nl}\,=\,\sqrt{-2\,c'\,D}\{\frac{\psi_+}{2}\,-\,\psi_-\}. \label{schvnl}
\end{equation}

Given the results found for the previous model, we expect that our system 
(\ref{schlgleffectivepot}) will present fronts in the linear regime for a 
high enough noise intensity $\epsilon(0)$ and small enough $\tau$. 

Although we do not know the corrections of $d'$ beyond $O(\tau)$, 
 we expect the velocity to start diverging from (\ref{GLvUCNA})
for some finite value of $\tau$. But, by adequately choosing the parameters, 
the transition between the linear and nonlinear regimes can be
obtained at small enough values of $\tau$ for our approximations to
be applicable.
Hence, for values of the $a'$ near threshold, and
within a neighborhood of the crossing point, the nonlinear
velocity $v_{nl}$ will approximately be given by Eq.(\ref{schvnl}).

\begin{figure}
\centerline{\psfig{figure=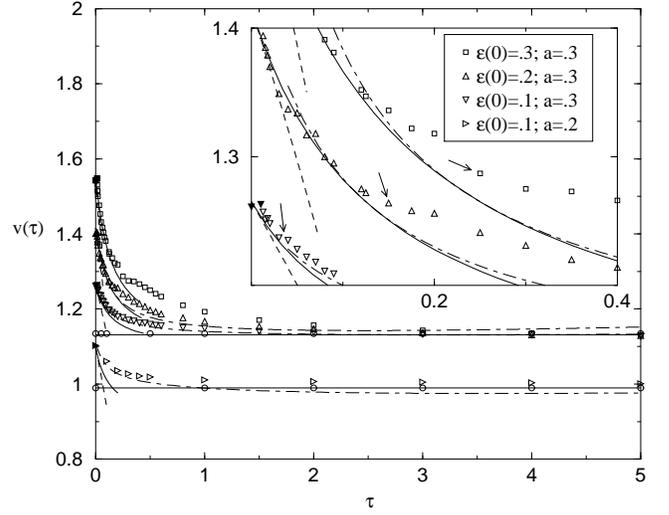,width=8.5cm}}
\caption{\em
Front mean velocity versus noise corelation time $\tau$ for the 
Schl\"ogl 
model.
Circles correspond to the deterministic
case, $v^*_{d,sim}=1.131$; dashed lines are the analytical results up 
to $O(\tau)$, whereas continuous lines represent the analytical 
corrected  results $v^*_{\tau}$ Eq.(\protect\ref{GLvUCNA}).
Dot-dashed lines correspond to the prediction (\protect\ref{schvnl}).
The inset amplifies the domain of small $\tau$ .
By tuning $\tau$, the front shifts from a {\sl pulled} regime to a
{\sl pushed} one. This crossover corresponds to the points first leaving
the theoretical curve $v^*_{\tau}$ and are approximately given by the
arrows.
 \label{schvtau}
}
\end{figure}

We will show that numerical simulations support this analysis.  
In Fig.(\ref{schvtau}) we have plotted the numerical results
of the front mean velocity versus $\tau$. 
For all three plots, the value of the deterministic linear coefficient
$a$ is such that the deterministic front lies right within the
nonlinear regime. While for the white noise case  
all fronts move within the linear regime, the one for 
$\epsilon(0)=.1$ is only marginally inside. 
Increasing then $\tau$, the effective linear coefficient $a'$ decreases.
For small values of the correlation time, the fronts will still lie
within the linear regime, except for the $\epsilon(0)=.1$ case, for
which the front enters the nonlinear regime immediately for
any finite value of $\tau$.

Thus for small values of $\tau$ the front moves with the linear velocity.
We can see that our first order approximation (dashed lines) also
reproduces for this system the initial steep fall of $v$. Our analytical
continuation $v^*_{\tau}$ shows up a perfect agreement with the numerical
results.
By further increasing the correlation time we can shift the front into
the non-linear regime. 

For this one we only have a rough approximation
for the front mean velocity given by (\ref{schvnl}) (dot-dashed lines).
This approximation is valid only near the critical point $a'=0$ and
gets worse as we move away from it. Although our analysis gives us
only the corrections due to a finite $\tau$ up to {\sl O}($\tau$),
with the renormalized noise intensity (\ref{e_R}), and staying close
to the critical point, we obtain a surprisingly good analytical estimation
for the non linear velocity up to values of $\tau$ of order {\sl O}($1$).
One can notice that our predictions are just {\sl O}($\tau$) seeing that
the velocity diverges from the deterministic value at $\tau$ high enough.
This  is to be expected as in obtaining (\ref{schvnl}) we have
used only an $O(\tau)$ approximation for the 
squared $b'\psi^2$ and the stabilizing term $c'\psi^3$,
with both, $b'$ and $c'$ linearly increasing with $\tau$ for high values of 
$\tau$.
 But this dependence
is obviously incorrect as a infinite energy difference between two
metastable states would give rise to an unbounded propagation velocity,
which is not the case.

The estimation of the crossing point between the linear and the non linear
regimes, given by the crossing point between the two 
curves $v^*_{nl}$ and $v^*_{\tau}$, is consistent with the numerical results.
At that point, these show up
a pronounced and increasing departure off the linear velocity $v^*_{\tau}$.
For the $\epsilon(0)=.1$ case, 
the departure from the theoretical result $v^*_{\tau}$ starts already for
$\tau>0$.
This can be best seen in the inset of figure \ref{schvtau}.

\section{Conclusions}
We have studied  a general reaction-diffusion system which
exhibits
fronts in the presence 
of spatio-temporal structured
external noise.
We have derived an 
 effective deterministic dynamical equation for the front 
which
contains the main effects of the noise. 
These show up as renormalization
of the original reaction terms of the deterministic system and also  new
terms.
The relevance of those effects are discussed for two prototype models:
The Ginzburg-Landau model (noise linear coupling) and the Schloegl model
(noise nonlinear coupling).
Although, our analysis is valid only for small correlation times, 
we are able to grasp  at least the main features
of the effects of a time colored noise on an extended system,
namely that there is a pronounced slow down of the front velocity already
for small but finite correlation times $\tau$. 

We have obtained an analytical expression for the
front velocity in the linear regime that we have extended to be valid for
any
value of the correlation time $\tau$ of the noise. The numerical
simulations present an excellent agreement with the analytical results in
this regime.
For the non linear regime,
we can only give an approximate expression
for the velocity, when the system is near threshold. Although it 
still is valid only for $\tau$ not too large, it goes beyond the 
first order approximation considered in our analysis when the system
is brought close enough to threshold.

Finally, we have obtained the non trivial influence of the noise
correlation length. 
Our results here are of a more limited validity, and this case 
needs a further study. Nevertheless, our numerical results 
suggest a non trivial behavior of the velocity when
varying the correlation length of the noise, with an increase
of the velocity with $\lambda$ for finite $\tau$ and small $\lambda$.
This would be a novel result, because previous reported studies
evaluate only the trivial dependence on $\lambda$ of the 
front velocity, slowing down  with increasing correlation length
\cite{ArJcpc}.

Hence, we have shown that our procedure of separating the systematic
contribution of the noise from the original dynamics gives reliable
information for front dynamics
in the presence of spatio-temporal structured noises. 
The  systematization of the present procedure, if possible at all, 
and its application to other situations would be extremely interesting. 

\section{ACKNOWLEDGMENTS}
We acknowledg financial support from the Ministerio de Ciencia
y Tecnolog\'ia (Spain) under project BFM2000-0624.
M.A.S acknowledges financial support from the Departament
d'Universitats, Recerca i Societat de la Informaci\'o, Generalitat
de Catalunya.
M.A.S is also very pleased to thank Prof. L. Schimansky-Geier for
fruitful comments and his hospitaly during a stay at the {\sl Institut
f\"{u}r Physik, Humbolt Universit\"{a}t zu Berlin}, where part of this
work was done.

\appendix

\section{Analytical derivation of the systematic effects of a structured 
noise}\label{rspfA}

The systematic
contribution $\langle \Phi (x,t) \rangle$ of the noise is given by
\begin{equation}
\langle\Phi (x,t)\rangle \equiv 
\epsilon^{1/2}\langle g\lgroup\psi(x,t)\rgroup \eta(x,t)\rangle.
\end{equation}
This average con be calculated by using Novikov's theorem in the following
form,
\begin{eqnarray}
\langle\Phi(x,t)\rangle = \epsilon^{1/2} \int_{\Re}dx'\int_0^tdt'\,
G\lgroup |x-x'|,|t'-t|\rgroup \nonumber \\
\,\bigg< g'\lgroup \psi(x,t)\rgroup 
\frac{\delta \psi(x,t)}{\delta \eta(x',t')} \bigg>. \label{Phi}
\end{eqnarray}
Hence, the determination of $\langle\Phi(x,t)\rangle$ reduces to that 
of the response function
\begin{equation}
Q(x,x';t,t')\equiv \frac{\delta \psi(x,t)}{\delta \eta(x',t')}.
\label{rspfdef}
\end{equation}
Following \cite{SSKG} and \cite{LJm88}, we 
will consider the contribution of the noise at first order in $\tau$ in 
the approximation of small $\tau$. This means that temporal correlation 
decays very strongly for $t'\not= t$ 
leaving relevant in (\ref{rspf}) only the values of the integrant 
for $t'$ close to $t$. 
Thus we may expand $Q(x,x';t,t')$ in powers of $(t'-t)$ around $t'=t$ and 
take all up to the first order
\begin{equation}
Q(x,x';t,t')= Q(x,x';t,t)+\frac{\partial Q}{\partial t'}\Bigg|_{t'=t} 
(t'-t) + \cdots .\label{rspfexp}
\end{equation}
In appendix \ref{rspfB} we present a detailed derivation of this second term.

Now we can rearrange (\ref{Phi}) in two terms, the first one being 
the zero order or white noise contribution, i.e, 
the one we get in the limit $\tau\to 0$ for 
fixed $\epsilon$ and $\lambda$, while the second one represents 
the contribution of the colored noise at first order 
in $\tau$, 
\begin{equation}
\langle\Phi(x,t)\rangle=\langle\Phi_0(x,t)\rangle\,+\,
\langle\Phi_1(x,t)\rangle. \label{Phidec}
\end{equation}
Recollecting relations (\ref{Phi}),(\ref{rspfexp}) and 
(\ref{rspfic}),(\ref{rspfdrv2}), and after calculating the spatial integral,
 we obtain
\begin{eqnarray}
&&\langle\Phi_0(x,t)\rangle =
\epsilon\,\Big<g\lgroup\psi(x,t)\rgroup\,g'\lgroup\psi(x,t)\rgroup\Big>
\int_0^tdt'\,G\lgroup0,|t-t'|\rgroup\label{Phi0t} \nonumber \\
&&\langle\Phi_1(x,t)\rangle =
\epsilon\,
\Big<
\,g'\lgroup\psi(x,t)\rgroup\,
\Bigg[\Big[
\{ g\lgroup\psi(x,t)\rgroup,f\lgroup\psi(x,t)\rgroup\}\,  \\
&&-\,D\,g''\lgroup\psi(x,t)\rgroup\,
\Big(\frac{\partial\psi(x,t)}{\partial x}\Big)^2\, 
\Big]
\,\int_0^tdt'\,G\lgroup0,|t-t'|\rgroup\,(t'-t)\,\nonumber \\
&&-\,D\,g\lgroup\psi(x,t)\rgroup
\int_0^tdt'\,G''\lgroup0,|t-t'|\rgroup\,(t'-t)
\Bigg]
\Big>\,\nonumber \label{Phi1t}
\end{eqnarray}
where the primes on $G\lgroup0,(t-t')\rgroup$ indicate 
derivatives of $G\lgroup(x-x'),(t-t')\rgroup\,$ 
with respect to $x'$, evaluated at $x'=x$, and 
\begin{eqnarray}
\{ g\lgroup\psi(x,t)\rgroup,f\lgroup\psi(x,t)\rgroup \}\,&\equiv\,&
g'\lgroup\psi(x,t)\rgroup\,f\lgroup\psi(x,t),a\rgroup\,-\,\nonumber \\
&-&g\lgroup\psi(x,t)\rgroup\,f'\lgroup\psi(x,t),a\rgroup.
\label{poissonbra} 
\end{eqnarray}
As we are interested in the approximation of small $\tau$,  
which amounts to consider observation times much more greater 
than the characteristic correlation time 
of the noise, we can extend then the limits of these 
integral up to $\infty$. 

At this point, further assumptions
on the correlation function must be done in order to obtain
any analytical prediction. Assuming that $G(x,s)$ 
factorizes as in (\ref{etastat}),
the above integrals can be written as
\begin{eqnarray}
\,\int_0^{t}dt'\,G\lgroup0,|t-t'|\rgroup\,&=&\,
C(0) \sim \frac{\sigma^2}{\lambda} \nonumber \\
\,\int_0^{t}dt'\,G\lgroup0,|t-t'|\rgroup\,(t'-t)\,&=&\,
-\,\tau\,C(0) \sim \frac{\sigma^2 \tau}{\lambda}\nonumber \\
\,\int_0^{t}dt'\,G''\lgroup0,|t-t'|\rgroup\,(t'-t)\,&=&\,
-\,\tau\,C''(0) \sim \frac{\sigma^2 \tau}{\lambda^3}
. \label{gpartial}
\end{eqnarray} 
where the temporal part $\gamma(s)$ is considered to be
normalized to $1$.

Finally, we can write
\begin{eqnarray}
\langle\Phi_0(x,t)\rangle&=&
\epsilon\,C(0)\, 
\Big<g\lgroup\psi(x,t)\rgroup\,g'\lgroup\psi(x,t)\rgroup\Big>
\label{Phi0} \\ 
\langle\Phi_1(x,t)\rangle&=&
-\,\epsilon\,C(0)\,\tau\,
\Big<
\,g'\lgroup\psi(x,t)\rgroup\,
\Big[
\{ g\lgroup\psi(x,t)\rgroup,f\lgroup\psi(x,t)\rgroup\} \nonumber\\ 
&-&\,D\,g''\lgroup\psi(x,t)\rgroup\,
\Big(\frac{\partial\psi(x,t)}{\partial x}\Big)^2\, 
\Big]
\Big>\nonumber \\
&+&\,\epsilon\,D\,C''(0)\,\tau\,
\Big<
g'\lgroup\psi(x,t)\rgroup\,
g\lgroup\psi(x,t)\rgroup
\Big>. \label{Phi1}
\end{eqnarray}

\section{Response function}\label{rspfB}
The determination of $\langle\Phi(x,t)\rangle$ reduces to that 
of the response function
\begin{equation}
Q(x,x';t,t')\equiv \frac{\delta \psi(x,t)}{\delta \eta(x',t')}.
\label{Arspfdef}
\end{equation}
The meaning of (\ref{SPDE}) is given by
\begin{eqnarray}
\psi(x,t+\Delta t)&-&\psi(x,t)=\int_{t}^{t+\Delta t} ds\, 
{\cal L} \lgroup \psi(x,s),\partial_x,a\rgroup \nonumber \\
&+&\epsilon^{1/2} \int_t^{t+\Delta t} ds\, g\lgroup\psi(x,s)\rgroup
\,\eta(x,s), \label{meaningSPDE}
\end{eqnarray}
and as long as $\tau>0$, the last integral is well defined as a Riemann 
integral.
By taking the functional derivate of (\ref{meaningSPDE}) with 
respect to the noise $\eta(x',t')$ we get 
\begin{eqnarray}
Q(x,x';t,t') & = & 
Q(x,x';t',t')  
+\nonumber \\
\int_{t'}^tds\,
\{
\frac{\partial {\cal L}}{\partial\psi }(x,s)\,&+&\,
\epsilon^{1/2} g'\lgroup \psi(x,s)\rgroup \,\eta (x,s) 
\}  
\, Q(x,x';s,t')\nonumber\\
&;&\,\,t\,>\,t'\label{rspf}
\end{eqnarray}
with
\begin{equation}
Q(x,x';t,t)=\epsilon^{1/2}\,g\lgroup\psi(x,t)\rgroup\,
\delta(x'-x).\label{rspfic}
\end{equation}
Equation (\ref{rspf}) is an integro-differential equation for the 
response function for 
which it has not yet been  found  a formal
solution 
as it was  in \cite{SSKG} and \cite{LJm88} for non spatially
dependent 
(0-dimensional) systems. The term (\ref{rspfic}) gives the contribution
(\ref{Phi0}) to the systematic effect of the noise.
Expanding $Q(x,x';t,t')$ in powers of $(t'-t)$ around $t'=t$ and 
taking all up to the first order
\begin{equation}
Q(x,x';t,t')= Q(x,x';t,t)+\frac{\partial Q}{\partial t'}\Bigg|_{t'=t} 
(t'-t) + \cdots .\label{Arspfexp}
\end{equation}
The second term can be obtained by directly deriving (\ref{rspf}) 
with respect to $t'$. This gives
\begin{equation}
\frac{\partial Q(x,x';t,t')}{\partial t'}\,=\,
\int_{t'}^tds {\cal I}(x,s) 
-\{\,\frac{\partial\psi(x,t')}{\partial t'}\,,\,Q(x,x';t,t')\}
\label{Arspfdrv1}
\end{equation}
As long as we are interested in the limit $t'\to t$, the details 
of ${\cal I}(x,s)$ are not important for the last term in 
(\ref{Arspfdrv1}) vanishes in that limit since it is a regular function in $s$.
Fort $t'=t$, and substituting equation (\ref{SPDE}), 
\begin{eqnarray}
\frac{\partial Q(x,x';t,t')}{\partial t'}\Bigg|_{t'=t}&=&
\frac{\partial Q(x,x';t',t')}{\partial \psi(x,t')}\Big|_{t'=t}\,\nonumber \\ 
\Bigg( {\cal L}(\psi (x,t),\partial_x ,a)\,&+&
\,\epsilon^{1/2}\,g\lgroup\psi(x,t)\rgroup\,\eta(x,t)
\Bigg)\,-\, \nonumber \\
-\Bigg( \frac{\partial {\cal L}}{\partial\psi }(x,t)\,+
&\epsilon^{1/2}&
g'\lgroup\psi(x,t)\rgroup\,\eta(x,t) 
\Bigg) Q(x,x';t,t)\,. \label{rspfdrv2} 
\end{eqnarray}
Considering the initial condition (\ref{rspfic}) and
substituting the expression of the
nonlinear (differential) operator ${\cal L}$ given by 
(\ref{SPDE}), the last relation reduces to
\begin{eqnarray}
\frac{\partial Q(x,x';t,t')}{\partial t'}\Bigg|_{t'=t}=
\epsilon^{1/2}\,
\Big[\,g'\lgroup\psi(x,t)\rgroup\,f\lgroup\psi(x,t),a\rgroup\,&-&\,\nonumber \\
g\lgroup\psi(x,t)\rgroup\,f'\lgroup\psi(x,t),a\rgroup
\Big]\delta (x'-x)\,&+&  \nonumber \\
D\,\epsilon^{1/2} \,
\{
g'\lgroup\psi(x,t)\rgroup\,
\lgroup\frac{\partial^2\psi(x,t)}{\partial x^2}\rgroup\,\delta (x'-x)\,&-&\,
\nonumber \\
\frac{\partial^2}{\partial x^2}\Big[g\lgroup\psi(x,t)\rgroup\,
\delta (x'-x)\Big]
\}.&&  \label{Arspfdrv4} 
\end{eqnarray}
The second order derivative of the last term gives
\begin{eqnarray}
\Big[\,g''\lgroup\psi(x,t)\rgroup\,
\lgroup\frac{\partial\psi(x,t)}{\partial x}\rgroup^2\,+\,
g'\lgroup\psi(x,t)\rgroup\,
\lgroup\frac{\partial^2\psi(x,t)}{\partial x^2}\rgroup\Big]\,
\nonumber \\
 \delta (x'-x)\,+\,2\,g'\lgroup\psi(x,t)\rgroup\,
\lgroup\frac{\partial\psi(x,t)}{\partial x}\rgroup\,
\frac{\partial\delta (x'-x)}{\partial x}\, \nonumber \\
+\,g\lgroup\psi(x,t)\rgroup\,
\frac{\partial^2\delta (x'-x)}{\partial x^2}. \label{Adrvdelta}
\end{eqnarray}
With this result in mind, the Laplacian terms in (\ref{Arspfdrv4})
mutually cancel, while the terms proportional to the $\delta$ give
rise to the first integral in (\ref{Phi1}).
Taken into account the following relation of the derivative of a $\delta$
\begin{equation}
\frac{\partial}{\partial x}\delta(x'-x)\,=\,-\,
\frac{\partial}{\partial x'}\delta(x'-x) 
\end{equation}
the contribution to (\ref{Phi}) of the first and second order 
derivative of the $\delta$ in (\ref{Adrvdelta}) will be
\begin{eqnarray}
-\int_0^tdt'\,
\Big[2\,G'\lgroup0,(t-t')\rgroup\,g'\lgroup\psi(x,t)\rgroup
\,\frac{\partial \psi(x,t)}{\partial x}\,+\,\nonumber \\
G''\lgroup0,(t-t')\rgroup\,
g\lgroup\psi(x,t)\rgroup\Big]\,(t'-t).
\end{eqnarray}

This result give rise to the last terms
in (\ref{Phi1}),
where the contribution proportional to 
$G'\lgroup 0,(t-t')\rgroup$
has been discarded because of the spatial isotropy of the noise.

\section{Numerical algorithm for generating a spatio-temporal colored
noise}\label{stcn}

Here, we will define a spatio-temporal
structured noise that is very simple to implement numerically, and
which is the one we have used in this work. This type of noise
is obtained
by rewriting the spectral method \cite{JgoJmbook} as a linear
transformation
of a more simple noise field in real space.

We define our spatio-temporal colored noise in each lattice cell $i$ and 
at time $t$, as
\begin{equation}
\eta_i(t)\,\equiv\,\Delta x\,\sum_i\,\bar{\eta}_j(t)\,g_{i-j} \label{cnvn}
\end{equation}
where the index $j$ labels a domain of cells around the cell $i$, and 
$g_i$ is a weighting distribution with the isotropic property
\begin{equation}
g_{-i} \,=\, g_{i}. \label{gsymetry}
\end{equation}

$\bar{\eta}_i(t)$ is an Ornstein-Uhlenbeck process in the lattice cell $i$, 
statistically independent of the other lattice points (white noise in
space). Its value is generated through the linear
Langevin equation,
\begin{equation}
\frac{\partial \bar{\eta}_i(t)}{\partial t} =
-\frac{1}{\tau}\bar{\eta}_i(t)+\frac{1}{\tau} \xi_i(t),
\end{equation}
in terms of a Gaussian white noise with a correlation,
\begin{equation}
\langle \xi_i(t') \xi_j(t) \rangle \,=\, 2
\,\sigma^2_w\,\frac{\delta_{i,j}}
{\Delta x} \delta(t-t').
\end{equation} 
In this case the correlation of the noise $\bar{\eta}_i(t)$, is given by
\begin{eqnarray}
\langle \bar{\eta}_{l+j}(s)\bar{\eta}_j(0) \rangle = \bar{G}_l(s) &=&
\bar{C}_l\,\gamma (|s|). \nonumber \\
\gamma (|s|) &=& \frac{\sigma^2_w}{\tau}\,e^{-\frac{|s|}{\tau}}
\label{OU-g}
\\
\bar{C}_l&=&\frac{\delta_{0,l}}{\Delta x}. \nonumber
\end{eqnarray}
Being $\bar{G}$ already factorized, the linear transformation (\ref{cnvn})
assures that the correlation function of $\eta_i(t)$ will be of the
desired form
(\ref{etastat}).
As $g$ is arbitrary, we are free to impose the condition that the value
of $G_0(0)$ equals that of $\bar{G}_0(0)$, i.e,
\begin{equation}
G_0(0)\,=\,\bar{G}_0(0) \label{gspatial_norm}
\end{equation}
Here we are  interested in $g_l$ having a finite range, for simplicity,
we assume that $g_i$ is a constant $g$  inside the
interval
$-m\,<\,i\,<\,m$, but zero otherwise. Then the  condition
(\ref{gspatial_norm}) implies $g_i\,=\,
(\,\Delta x\,\sqrt{(2\,m\,+\,1)}\,)^{-1}$.
Now it is a simple calculation to show that $\eta_i(t)$ is a
spatio-temporal 
structured noise with a correlation,
\begin{eqnarray}
G_l\,=\,C_l\,\gamma(|s|)
= \frac{\bar{C}_0}{(2\,m\,+\,1)}\,\left[2\,m\,+\,1\,-\,|l|\right]\,
\nonumber \\
\theta(2m-|l|) \gamma(|s|).  
\end{eqnarray}

At equal lattice points this function decays exponentially in time, and at
equal times it has a triangular decay as a function of lattice point
difference. From this analytical expression it is straightforward to obtain
the noise 
intensity and correlation length, 
\begin{eqnarray}
\sigma^2 = (2m+1)\sigma^2_w
\nonumber\\
\lambda=\sqrt{\frac{2}{3}\,m\,(m\,+\,1)}\,\Delta x, 
\label{definitions}
\end{eqnarray}
being the correlation time $\tau$.
One can check now that in the lattice white noise limit  $m=0$
and $\sigma^2\,=\,\sigma^2_w$, and then $\lambda\,=\,0$. For $m=1$, we
get that $\lambda\,=1.15...\,\Delta x$.

%

\end{document}